\documentclass[a4paper]{article}
\usepackage{interspeech2013,amssymb,amsmath,epsfig}
\usepackage{stfloats}
\ninept
\def\reg{{\rm\ooalign{\hfil
     \raise.07ex\hbox{\scriptsize R}\hfil\crcr\mathhexbox20D}}}

\title{Exploring Deep Hybrid Tensor-to-Vector Network Architectures for Regression Based Speech Enhancement}

\makeatletter
\def\name#1{\gdef\@name{#1\\}}
\makeatother

\name{Jun Qi$^{1}$, Hu Hu$^{1}$,
 Yannan Wang$^{3}$, Chao-Han Huck Yang$^{1}$, Sabato Marco Siniscalchi$^{1,2}$, Chin-Hui Lee$^{1}$ }
\address{$^1$Electrical and Computer Engineering, Georgia Institute of Technology, Atlanta, GA, USA \\
$^2$Computer Engineering School, University of Enna, Italy \\
$^3$Tencent Media Lab, Tencent Corporation, Shenzhen, Guangdong, China}

\begin{document}
\maketitle

\begin{abstract}
This paper investigates different trade-offs between the number of model parameters and enhanced speech qualities by employing several deep tensor-to-vector regression models for speech enhancement. We find that a hybrid architecture, namely CNN-TT, is capable of maintaining a good quality performance with a reduced model parameter size. CNN-TT is composed of several convolutional layers at the bottom for feature extraction to improve speech quality and a tensor-train (TT) output layer on the top to reduce model parameters. We first derive a new upper bound on the generalization power of the convolutional neural network (CNN) based vector-to-vector regression models. Then, we provide experimental evidence on the Edinburgh noisy speech corpus to demonstrate that, in single-channel speech enhancement, CNN outperforms DNN at the expense of a small increment of model sizes. Besides, CNN-TT slightly outperforms the CNN counterpart by utilizing only 32\% of the CNN model parameters. Besides, further performance improvement can be attained if the number of CNN-TT parameters is increased to 44\% of the CNN model size. Finally, our experiments of multi-channel speech enhancement on a simulated noisy WSJ0 corpus demonstrate that our proposed hybrid CNN-TT architecture achieves better results than both DNN and CNN models in terms of better-enhanced speech qualities and smaller parameter sizes.

\end{abstract}

\section{Introduction}
\label{sec:intro}

A speech enhancement system aims at restoring the quality and intelligibility of noisy speech. The state-of-the-art speech enhancement systems are commonly built with deep neural network (DNN) based vector-to-vector regression models, where inputs are context-dependent log power spectrum (LPS) features of noisy speech and outputs correspond to either clean or enhanced LPS features. Although deep neural network (DNN) based speech enhancement \cite{xu2015regression, dwang2018} has demonstrated the state-of-the-art performance under a single-channel setting, it can also be extended to scenarios of multi-channel speech enhancement with even better-enhanced speech qualities \cite{wang2018two}. The process of both single and multi-channel speech enhancement can be taken as a DNN based vector-to-vector regression aiming at bridging a functional relationship $f: \mathbb{Y} \rightarrow \mathbb{X}$ such that the input noisy speech $ y \in \mathbb{Y}$ can be mapped to the corresponding clean speech $ x \in \mathbb{X}$. In~\cite{xu2015regression, yu2020time}, DNNs with feed-forward fully-connected (FC) hidden layers were proposed to attain the state-of-the-art performance of speech enhancement on the target tasks and the related theorems were later set up in~\cite{qi2019theory, qi2020theory, mae_spl}. In some follow-up studies, recurrent neural networks (RNNs)~\cite{weninger2015speech, zhao2018convolutional}, and convolutional neural networks (CNNs)~\cite{park2016fully} were further investigated to boost speech enhancement quality~\cite{yang2020characterizing}. Moreover, a deep bidirectional RNN with LSTM gates was instead used in~\cite{sun2015voice}, and a generative adversarial network (GAN) was attempted for speech enhancement tasks in~\cite{pascual2017segan}. In particular, CNN is a tensor-to-vector regression model because it is capable of dealing with 3D/4D tensorized input data. Besides, the recent works~\cite{park2016fully, kounovsky2017single} suggest that CNN can outperform both DNN and RNN counterparts for speech enhancement. Similarly, a tensor-to-vector regression model can also be built by directly employing the proposed tensor-train network (TTN)~\cite{novikov2015tensorizing}. Besides, TT-DNN is a compact representation for a fully-connected (FC) layers of DNN into a tensor-train (TT) format~\cite{oseledets2011tensor}. In~\cite{jun2020}, we were the first to attempt a tensor-train deep neural network (TT-DNN) to tackle the multi-channel speech enhancement task and also demonstrate that the TT representation of a DNN does not cause the quality degradation of the enhanced speech, and it also results in a significant reduction of the model parameters. More importantly, the quality of speech enhancement can be improved over the DNN counterpart by allowing the TT-DNN parameters to grow.

A significant advantage of tensor-to-vector regression, such as CNN and TT-DNN, is its compact architecture to observe stringent hardware constraints, where computational resources are often limited. Therefore, it is worth investigating the models in terms of the representation power, and experimentally comparing them by considering the trade-off between enhancement performance and the number of model parameters. On one hand, CNN is a powerful model to learn spatial-temporal features and extract semantically meaningful aspects in higher hidden layers. On the other hand, TT-DNN can maintain baseline results of the corresponding DNN by applying the TT transformation to the FC hidden layers. Hence, in this work, we focus on a tensor-to-vector model to take advantage of both CNN and TT-DNN. More specifically, we propose a novel hybrid architecture, namely CNN-TT, with convolutional layers stacked at the bottom and one TT hidden layer on the top. To highlight the advantages of CNN-TT, we compare different deep tensor-to-vector models for speech enhancement. The used models in this work include (a) DNN; (b) CNN; (c) TT-DNN; (d) CNN-TT. In more detail, we first explain the fundamental mechanisms of tensor-to-vector regression based on our theorems of DNN based vector-to-vector regression~\cite{qi2019theory, qi2020theory, qi2021tensor, qi2020performance}. Then, we validate our CNN-TT models in speech enhancement tasks.

Our experimental results show that in single-channel speech enhancement on the Edinburgh noisy speech corpus \cite{valentini2016investigating}, CNN outperforms the best DNN with a small increment of parameter sizes. Moreover, our proposed CNN-TT slightly outperforms CNN with only 32\% of the CNN model size. A further improvement can be attained if the size of the CNN-TT model is increased up to 44\% of the CNN model size. Finally, the experiments of a multi-channel speech enhancement task on a simulated noisy WSJ0 corpus \cite{Paul1992} show the same trend that our proposed hybrid CNN-TT architecture can be favorably compared to both DNN and CNN models to achieve better-enhanced speech qualities and utilize much smaller model sizes.

\section{Deep Tensor-to-vector Regression}
\label{sec:ttm}
\begin{figure*}[ht]
\centering
\centerline{\epsfig{figure=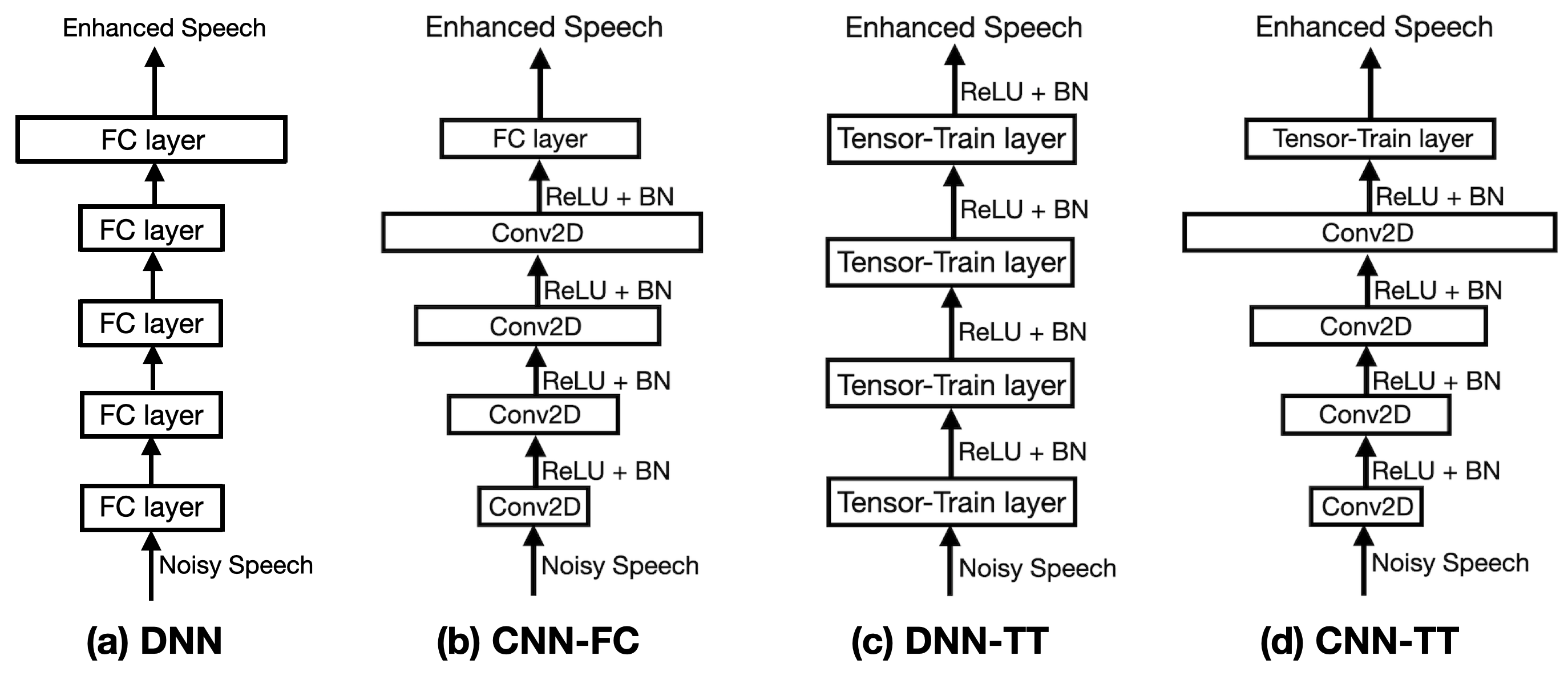, width=125mm}}
\caption{{\it Four tensor-to-vector regression models used in this study. }}
\label{fig:1}
\vspace{-3mm}
\end{figure*}

Figure~\ref{fig:1} shows all regression network architectures studied here: (a) DNN, (b) CNN, (c) DNN-TT, and (d) CNN-TT.

\subsection{CNN Based Tensor-to-vector Regression}
\label{sec:21}
CNN follows a feed-forward architecture to transform a tensor input into a vector output through a sequence of convolutional neural layers~\cite{garipov2016ultimate}. The CNN based tensor-to-vector regression model has four two-dimensional (2D) convolutional layers, each having twice the number of channels of the previous layer. ReLU-based activation and Batch normalization components are appended at the output of each convolutional layer. A fully-connected (FC) layer is employed as the last hidden layer of the neural architecture to generate the desired enhanced speech vectors.

A typical convolutional layer transforms a 3-dimension input tensor $\mathcal{X} \in \mathbb{R}^{W \times H \times C}$ into an output tensor $\mathcal{Y} \in \mathbb{R}^{(W-L+1) \times (H-L+1) \times S}$ by convolving $\mathcal{X}$ with a kernel tensor $\mathcal{K}\in \mathbb{R}^{L \times L \times C \times S}$ as:
\begin{equation*}
\mathcal{Y}(x, y, s) = \sum\limits_{i=1}^{L} \sum\limits_{j=1}^{l}\sum\limits_{c=1}^{C} \mathcal{K}(i, j, c, s) \mathcal{X}(x + i - 1, y + j - 1, c).
\end{equation*}

In \cite{qi2019theory}, we studied the representation power of DNN based vector-to-vector regression and derived upper bounds on different DNN architectures. That study allows us to better understand the successful application of DNN for speech enhancement tasks observed in \cite{xu2013experimental}. To extend the theorems proposed in \cite{qi2019theory} to CNNs, we need to obtain a matrix representation for both input and kernel of the CNN. Thus, we introduce a matrix $\textbf{X}$ of size $W'H' \times L^{2}C$, in which the $k$-th row corresponds to the $L \times L \times C$ patch of the input tensor that is used to compute the $k$-th row of the matrix $\textbf{Y}$:
\begin{equation*}
\begin{split}
&\hspace{4mm} \mathcal{X}(x + i - 1, y + j - 1, c) \\
&= \textbf{X}(x + W'(y-1), i + L(j-1) + L^{2}(c-1)),
\end{split}
\end{equation*}
where $y=1,...,H'$, $x=1,...,W', i, j = 1, ..., L$.

The kernel tensor $\mathcal{K}$ can be reshaped into a matrix $\textbf{K}$ of the size $l^{2} C \times S$ as follows:
\begin{equation*}
\mathcal{K}(i, j, c, s) = \textbf{K}(i + L(j-1) + L^{2}(c -1), s).
\end{equation*}

\begin{figure}[htbp]
\centerline{\epsfig{figure=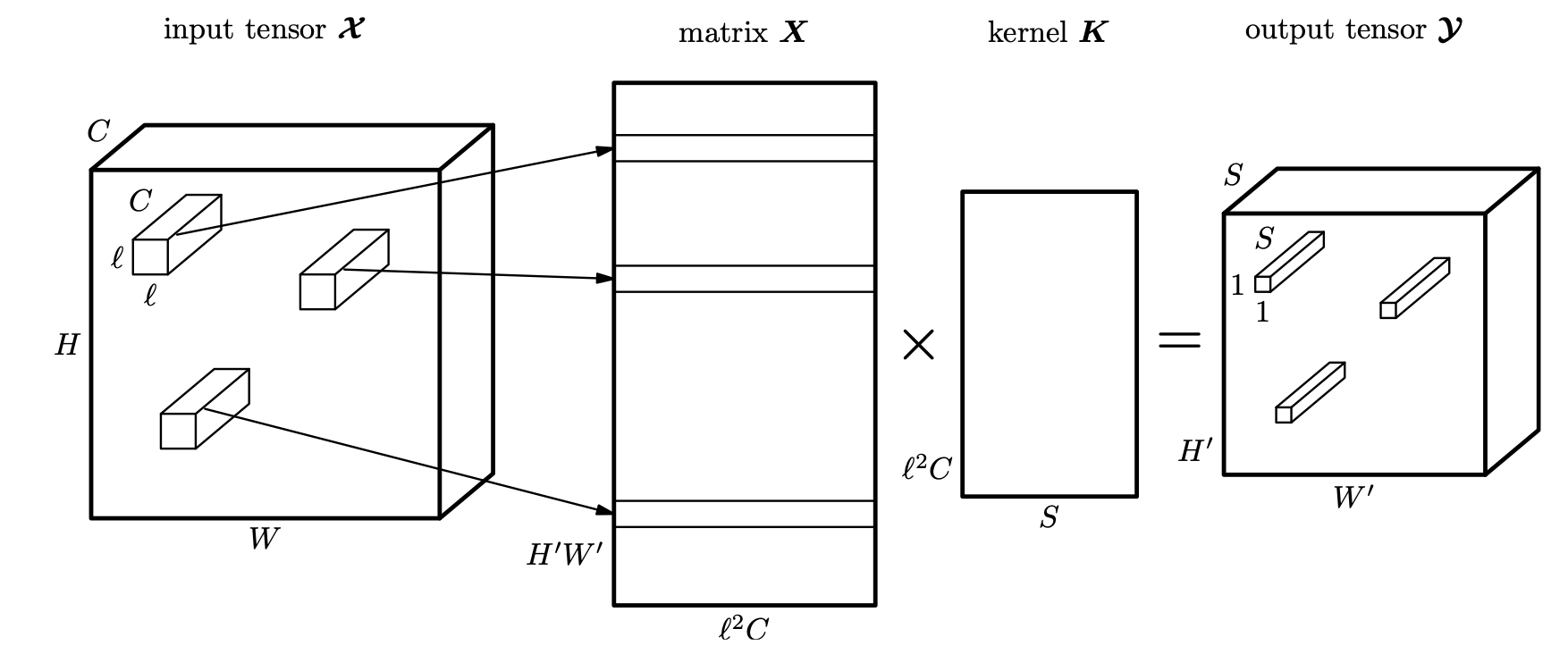, width=80mm}}
\caption{{\it Convolution as a matrix-by-matrix multiplication.}}
\label{fig:2}
\vspace{-3mm}
\end{figure}

Finally, a convolutional layer can be rewritten in a matrix format as $\textbf{Y} = \textbf{X} \textbf{K}$, and the process is illustrated as Figure~\ref{fig:2}.

We are now ready to link CNNs with our theorems for DNN-based vector-to-vector regression in \cite{qi2019theory}. Let $\hat{f}: \mathbb{R}^{d} \rightarrow \mathbb{R}^{q}$ refer to a vector-to-vector smooth function, we can find a deep CNN $f_{CNN}$ with $B$ layers with ReLU activations such that Eq. (\ref{eq:approx}) is satisfied,
\begin{equation}
\label{eq:approx}
\begin{split}
|| \hat{f} - f_{CNN} ||_{2} &\le || \hat{f} - f_{CNN} ||_{1} 	\\
 &= \mathcal{O}\left( \frac{q}{(L_{B}^{2}C_{B} + B - 1)^{\frac{1}{d}}} \right),
\end{split}
\end{equation}
where $C_{B}$ and $L_{B}$ denote the numbers of channel and width of the $B$-th CNN layer.

\subsection{DNN-TT Based Tensor-to-vector Regression}
\label{sec:22}
A DNN-TT based tensor-to-vector regression model relies on the TT decomposition, which is described as follows: For a set of integer ranks $\textbf{r} = \{r_{1}, r_{2}, ..., r_{K+1}\}$, the TT decomposition factorizes a tensor $\mathcal{W} \in \mathbb{R}^{(m_{1}n_{1})\times (m_{2}n_{2}) \times \cdot\cdot\cdot \times (m_{K}n_{K})}$, $\forall i\in \{1, ..., K\}, m_{i} \in \mathbb{R}^{+}, n_{i} \in \mathbb{R}^{+}$ into a multiplication of core tensors as:
\begin{equation}
\label{eq:ttn1}
\mathcal{W}((i_{1}, j_{1}), (i_{2}, j_{2}), ..., (i_{K}, j_{K})) = \prod\limits_{k=1}^{K} \mathcal{C}^{[k]}(r_{k}, i_{k}, j_{k}, r_{k+1}).
\end{equation}
where for the given ranks $r_{k}$ and $r_{k+1}$, the $k$-th core tensor $\mathcal{C}^{[k]}(r_{k}, i_{k}, j_{k}, r_{k+1}) \in \mathbb{R}^{m_{k} \times n_{k}}$ in which $i_{k} \in \{1, 2, ..., m_{K}\}$ and $j_{k} \in \{1, 2, ..., n_{k}\}$. Besides, $r_{1}$ and $r_{K+1}$ are fixed to $1$. Since DNN-TT only stores the low-rank core tensors $\{\mathcal{C}_{k}\}_{k=1}^{K}$ of the size $\sum_{k=1}^{K} m_{k}n_{k}r_{k}r_{k+1}$, which is much less than the size $\prod_{k=1}^{K}m_{k}n_{k}$ for the corresponding DNN.

\begin{figure}[t]
\centerline{\epsfig{figure=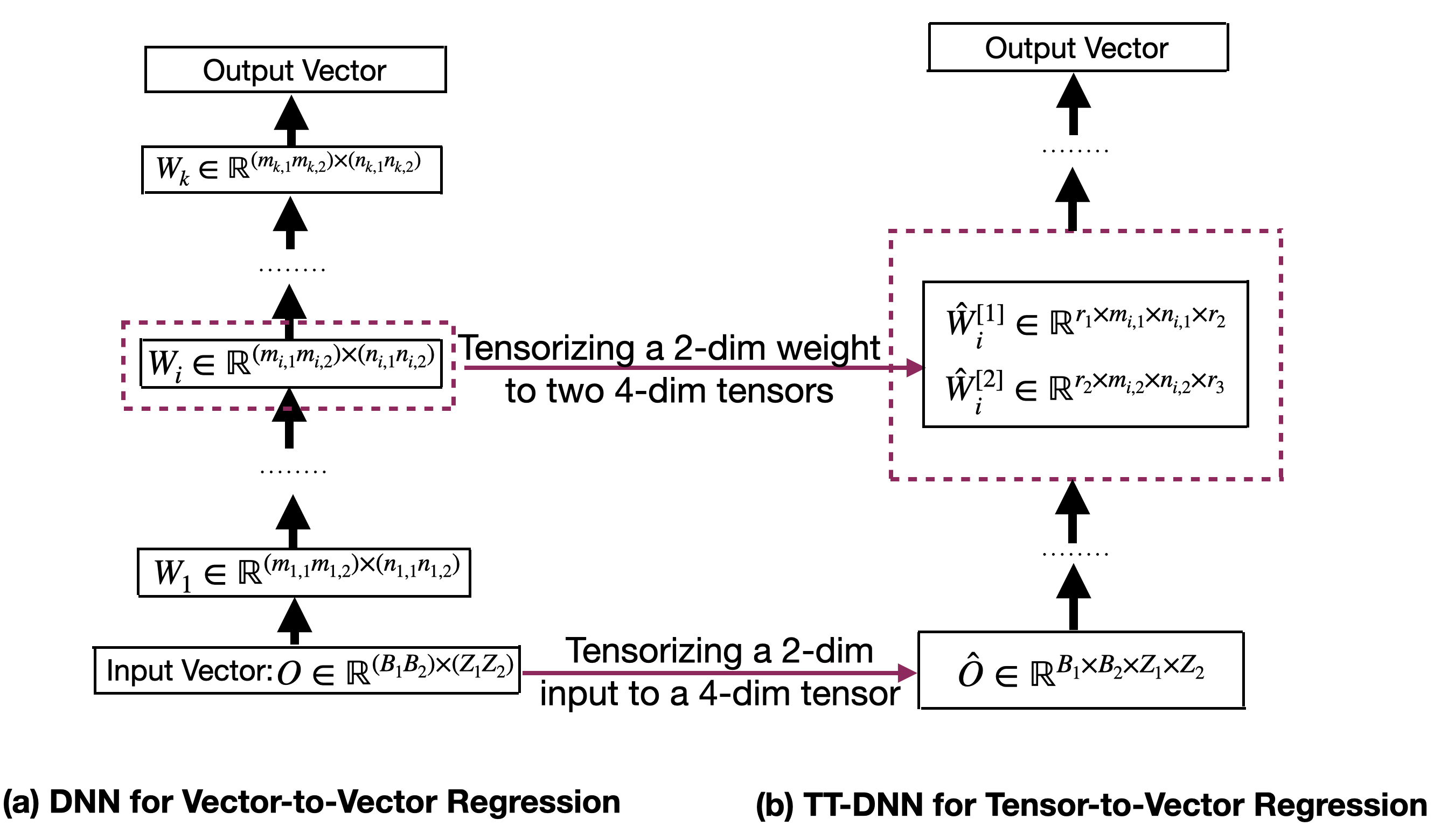, width=85mm}}
\vspace{-3mm}
\caption{{\it A conversion from a DNN to a DNN-TT. }}
\label{fig:3}
\vspace{-3mm}
\end{figure}

Figure~\ref{fig:3} shows the relationship between a traditional hidden layer of a DNN and a tensor layer of a DNN-TT. The matrix associated with a DNN hidden layer corresponds to two matrices given the ranks, and the DNN input vector is reshaped into a higher-order input tensor. We have shown that the TT decomposition can keep the representation power of DNN \cite{jun2020}. In \cite{jun2020}, we have also demonstrated that for a tensor-to-vector function $\mathcal{\hat{T}}^{*}: \mathbb{R}^{J_{1} \times J_{2}\times \cdot\cdot\cdot \times J_{K}} \rightarrow \mathbb{R}^{I_{1}\cdot I_{2}\cdot\cdot\cdot I_{K}}$, there is a DNN-TT $\mathcal{T}$ with $k$ hidden tensor layers, such that Eq. (\ref{eq:ttn2}) is satisfied.
\begin{equation}
\label{eq:ttn2}
\begin{split}
||\mathcal{\hat{T}} - \mathcal{T}||_{2} &\le ||\mathcal{\hat{T}} - \mathcal{T}||_{1}		\\
	&= \mathcal{O} \left( \prod\limits_{k=1}^{K} \frac{I_{k}}{{(r_{k-1}r_{l}n_{k, B}} + B - 1)^{\frac{1}{r_{k}r_{k-1}J_{k}}}} \right),
\end{split}
\end{equation}
where $n_{k, B}$ is the width of $B^{th}$ hidden layer for the $k$-th core tensor. The Eq.~(\ref{eq:ttn2}) suggests that DNN-TT can maintain the representation power of the corresponding DNN.

\subsection{CNN-TT Based Tensor-to-vector Regression}
Figure~\ref{fig:1}(c) displays a hybrid tensor-to-vector regression model having both convolutional and tensor-train layers. A key benefit of this hybrid tensorized model is that the number of model parameters of the original FC layer is significantly reduced with one TTN. Moreover, we can expect that the representation power of input salient features can be preserved because of the convolutional blocks in the lower layers.

The representation power of CNN-TT combines the characteristics of both CNN and TT-DNN, and Eq.~(\ref{eq:ttn3}) demonstrates an upper bound on the performance of CNN-TT based on tensor-to-vector regression. The derivation of the upper bound is based on the combination of Eqs.~(\ref{eq:ttn1}) and~(\ref{eq:ttn2}) ~\cite{jun2020}.
\begin{equation}
\label{eq:ttn3}
\begin{split}
||\mathcal{\hat{T}} - \mathcal{T}||_{2} &\le ||\mathcal{\hat{T}} - \mathcal{T}||_{1} 	\\
	&=\mathcal{O} \left( \prod\limits_{k=1}^{K} \frac{I_{k}}{{(r_{k-1}r_{l}c_{k, B}} + B - 1)^{\frac{1}{r_{k}r_{k-1}J_{k}}}} \right) .
\end{split}
\end{equation}
where $\prod_{k=1}^{K} c_{k, B} = L_{B}C_{B}$ and other notations are the same as Eqs.~(\ref{eq:ttn1}) and ~(\ref{eq:ttn2}).

\vspace{0.2cm}
\section{Experiments and Result Analysis}
\label{sec:experiments}
\subsection{Data Preparation}
The proposed architectures were evaluated on two different speech databases. One is based on the Edinburgh noisy speech database~\cite{valentini2016investigating}, where clean utterances were recorded from $56$ speakers including $28$ males and $28$ females from different accent regions both Scotland and the United States. Clean data were randomly split into $23075$ training and $824$ test waveforms, respectively. The noisy training speech materials, at four SNR levels: 15dB, 10dB, 5dB, and 0dB, were created from corrupting clean waveforms with the following noises: a domestic noise (inside a kitchen), and office noise (in a meeting room), three public space noises (cafeteria, restaurant, subway station), two transportation noises (car and metro), and a street noise (busy traffic intersection). In total, there were $40$ different noisy backgrounds for synthesizing the noisy training data (ten noises $\times$ four SNRs). As for the noisy test set, noise types included: a domestic noise (living room), an office noise (office space), one transport (bus), and two street noises (open area cafeteria and a public square). SNR values were: 17.5dB, 12.5dB, 7.5dB, and 2.5dB. Therefore, there were $20$ different noisy backgrounds for synthesizing the test data.

The second one is a synthesized database with 30-hour simulated materials obtained from the clean WSJ0 corpus \cite{Paul1992} with OSU-$100$-noise dataset~\cite{osu100}, which allows us to obtain $30$ hours of training waveforms and $5$ hours of test ones. To simulate the noisy data, each waveform was corrupted with one kind of background noise from the noise set. The target and additional interfering speech with their corresponding RIRs were convolved to generate the final noisy waveform. In doing so, the dataset contained additive noise, interfering speakers, and reverberation. 
 
Before we set up the training and testing sets, an improved image-source method (ISM) \cite{data_ism} was used to generate RIRs of reverberation time (RT$60$) (from $0.2s$ to $0.3s$) and the corresponding direct path response for each microphone channel. For both training and test datasets, the setting of RIRs was fixed to the same conditions, such as the room size, RT$60$, and all of the distances and directions. Additional detail about the data simulation procedure can be found in \cite{wang2018two, jun2020}.

\subsection{Experimental Setup}
In all experiments, we use $257$-dimensional normalized log-power spectral (LPS) feature vectors as inputs. LPS features were generated by computing $512$ points Fourier transform on a speech segment of $32$ milliseconds. For each input frame, $M$ neighboring adjacent frames were concatenated together, which results in a total $257 \times (2M + 1) \times B$ dimensional feature, where $B$ is the channel number of the input signal. As for the setup of TT-DNN, we ignored the first dimension of the input LPS features because it corresponded to the direct-current component. After the regression, the first dimension of input was concatenated back to the $256$-dimensional output without any change. The clean speech features were assigned to the top layers of tensor-to-vector regression models as the reference during the training stage.

The DNN based regression model was adopted as a baseline model. On the Edinburgh data set, the DNN model consisted of $4$ hidden layers with hidden dimensions configured to $1024$, $1024$, $1024$, $2048$, respectively. As for the WSJ0 simulated data set, we set up a $6$ layer DNN model with a hidden dimension of $2048$. Moreover, the CNN models kept similar deep tensor-to-vector structures in all experiments and were composed of four convolutional layers with gradually increasing the number of channels according to the setup of $32$-$64$-$128$-$128$. Moreover, the ReLU activation function and batch normalization were also utilized for each convolutional layer, and two FC layers with $2048$ neurons were stacked on the top layer to generate output vectors. Besides, we used different kernel sizes on the two datasets to obtain two slightly different model sizes.
Moreover, to improve the subjective perception in the speech enhancement tasks, the global variance equalization was applied to alleviate the problem of over-smoothing by correcting a global variance between estimated features and clean reference targets, and a technique of noise-aware training (NAT) was also employed to enable non-stationary awareness. Besides, the mean square error (MSE) loss was applied, which corresponds to the upper bounds of $L_{2}$ norm in Eqs. (\ref{eq:approx}), (\ref{eq:ttn2}), and (\ref{eq:ttn3}). Adam optimizer \cite{diederik2014} with an initial learning rate of $0.002$ was utilized during the training process, and the back-propagation (BP) algorithm was used to update the model parameters. The size of the context window at the input layer is set to $1$ for DNN in Edinburgh data, $5$ for DNN in WSJ0 simulated data, and $8$ for all CNN models. 
The perceptual evaluation of speech quality (PESQ)~\cite{rix2001perceptual}, was employed in our experimental validation.

\subsection{Single-channel Speech Enhancement Experiments}
\vspace{-1mm}
Table~\ref{tab:results} shows our experimental results on the Edinburgh noisy speech data set. Tensor-to-vector regression based on CNN can outperform the DNN baseline results in terms of a higher PESQ score ($3.03$ vs. $2.82$). DNN-TT with much fewer parameters ($0.55$M vs. $5.51$M) can maintain the same experimental performance of DNN, where the TT transformation was applied to the fully-connected layers. More importantly, compared with the combined convolutional and TT layers, the proposed CNN-TT can attain the highest PESQ score. If we allow the size of the CNN-TT model to increase up to 5.05M, a better speech enhancement quality can be attained with a PESQ score of 3.13.

\begin{table}[t]\footnotesize
    \caption{PESQ comparisons of single-channel deep speech enhancement models on the Edinburgh noisy speech database. The average PESQ score for unprocessed noisy speech is $1.97$.}
    \vspace{0.2cm}
    \centering
    \begin{tabular}{l|c|c}
    \hline
    \hline
    Model  		& Parameters \#   	& PESQ	 \\
    \hline
    \hline
    DNN		    &   5.5M		   & 2.82		\\ 
    \hline
    CNN    		    &  9.1M               	   & 3.04     	\\
    \hline
    DNN-TT	    &	0.55M		   & 2.81		\\
    \hline
    CNN-TT 	    & 0.73M	       	& 3.02		\\
    CNN-TT 	    &	2.9M	        	& 3.09		\\	
    CNN-TT 	    &   5.1M		 & 3.13		\\	
    \hline
    CNN-Tucker-3    & 8.9M               & 2.89		  \\
    \hline
    \hline
\end{tabular}
\label{tab:results}
\vspace{-3mm}
\end{table}

\subsection{Multi-channel Speech Enhancement Experiments}

The evaluation results on the 30-hour WSJ0 simulated multi-channel data are shown in Table~\ref{tab:results2}. The experimental results of both DNN and DNN-TT are in line with the results as shown in \cite{jun2020}. The usage of the DNN-TT model can significantly reduce the number of parameters without degrading the performance. Moreover, the CNN based tensor-to-vector regression model outperforms the DNN based one. Thus, CNN takes advantage of parameter reduction and the improvement of enhanced speech quality over DNN. In more detail, as for the single-channel case, CNN-TT attains a PESQ 3.04 using 2.8M parameters which correspond to CNN which attains the same PESQ score at 3.03 but costs more than 9.4M parameters. If the number of parameters is reduced to as small as 1.6M, the PESQ score is decreased to $2.99$. For our two-channel experiments, the CNN baseline has the same parameter numbers with a single channel one because the convolutional layer can be properly adapted to the multi-channel inputs. However, if the two fully connected layers in the CNN-based architecture with tensor-train layers, the model parameters can be significantly reduced from 9.4M to 2.8M without degrading the system performance in terms of the PESQ scores (3.13 vs. 3.11).

\begin{table}[t]\footnotesize
    \caption{PESQ comparisons of different deep models for multi-channel speech enhancement on the WSJ0 corpus. The average PESQ score for unprocessed ch-1 noisy speech is 2.02.}
    \vspace{0.2cm}
    \centering
    \begin{tabular}{l||c|c|c}
    \hline
    \hline
    Model  		&  Channel \# & Parameter \# & PESQ	 \\
    \hline
    \hline
    DNN		&	1		     & 	27M	        & 	2.86	\\

    DNN		&	2		     & 	33M	        & 	3.00	\\
    \hline
    CNN    	&   1		 &	9.4M  	& 	3.03\\

    CNN    	&   2		 & 	9.4M   	& 	3.11\\	
    \hline
    CNN-TT 		&		1	    &	1.6M      	& 	2.99	\\	

    CNN-TT 		&   	1		&  	2.8M		& 	3.04	\\		

    CNN-TT 		&		2	    &	1.6M      	& 	3.08	\\	

    CNN-TT 		&   	2		&  	2.8M		& 	3.13	\\	
   	\hline
    CNN-Tucker-3    &       1       &  9.2M         & 2.63      \\
    CNN-Tucker-3    &       2       & 9.2M          & 2.56      \\
   	\hline
   	\hline
    \end{tabular}
    \label{tab:results2}
    \vspace{-3mm}
\end{table}

\subsection{Experimental Comparison with Tucker Decomposition}
Tucker decomposition~\cite{kim2007nonnegative} is a higher-order extension to the singular value decomposition obtained by computing the orthonormal spaces associated with the different modes of a tensor. It is also meaningful to verify whether tucker decomposition applied to each CNN convolutional layer can lead to the same parameter reduction with a small drop in the PESQ value. We refer to this Tucker-reduced CNN as CNN-Tucker. Particularly, CNN-Tucker-3 means that we apply Tucker decomposition to the first three CNN hidden layers except the top one. The related results by using CNN-Tucker-3 in Tables \ref{tab:results} and \ref{tab:results2} demonstrate that high-order singular value decomposition is not sufficient to obtain a smaller size deep tensor-to-vector regression model without sacrificing the speech quality. 

\section{Conclusion}
\label{sec:conclusion}
We compare several tensor-to-vector regression models for speech enhancement. These models include CNN, DNN-TT, and the hybrid models composed of convolutional and TT layers, namely CNN-TT. We first discuss the representation power by linking tensor-to-vector regression to our earlier theories on DNN based vector-to-vector regression. Next, we evaluate these models for single-channel speech enhancement on the Edinburgh noisy speech database. Finally, we conduct multi-channel speech enhancement on a synthesized WSJ noisy corpus. Our experimental results suggest that CNN can outperform both DNN-TT and DNN with smaller regression errors and higher PESQ scores. Moreover, when the fully-connected output layer of CNN is replaced with a TT layer to generate a hybrid regression network, we achieve even better performances by gradually increasing the model size of the TT layer. In future work, we will investigate different tensor representations to reduce the parameters of the hidden convolutional layers.

\newpage\clearpage
\bibliographystyle{IEEEtran}
\bibliography{bib/ref}

\end{document}